\documentclass[11pt,leqno,fleqn]{article}
\usepackage{lscape,amssymb, amsmath,verbatim,graphicx}
\usepackage{epsfig,subfigure,float}
\usepackage{morefloats}
\usepackage{enumerate}
\usepackage{subfig}
\usepackage{setspace}
\usepackage{caption}
\DeclareCaptionFont{singlespacing}{\setstretch{0.5}}
\usepackage{booktabs}

\usepackage    {amsmath}
\usepackage    {amssymb}
\usepackage    {epsf}
\usepackage    {natbib}
\bibpunct{(}{)}{,}{a}{}{;}
\usepackage    {amsthm}

\newfont{\popis}{cmcsc10}

\theoremstyle{remark}

\theoremstyle{definition}

 \newcommand{\carka}{\raise0.2em\hbox{,}}

\newcommand{\begeqO}{\begin{eqnarray*}}
\newcommand{\eneqO}{\end{eqnarray*}}
\newcommand{\begeq}{\begin{eqnarray}}
\newcommand{\eneq}{\end{eqnarray}}

\newcommand{\nin}{\noindent}

\newcommand{\Sup}{\mathop{\sup}\limits}
\newcommand{\Inf}{\mathop{\inf}\limits}

\newtheorem{Th}{Theorem}[section]

\newtheorem{Le}[Th]{Lemma}
\newtheorem{re}{Remark}[section]
\newtheorem{prop}[Th]{Proposition}

\newcommand{\pp}{,\ldots,}
\newcommand{\oh}{1/2}
\newcommand{\ofrac}[1]{\frac{1}{#1}}
\newcommand{\lr}[1]{\left({#1}\right)}

\newcommand{\g}{\gamma}

\newcommand{\pr}{\textbf{Proof:}~}
\newcommand{\prend}{\hfill $\Box$\\}
\newcommand{\prendwol}{\hfill $\Box$}

\newcommand\relphantom[1]{\mathrel{\phantom{#1}}} 

\newcommand{\eps}{\varepsilon}

\newcommand{\floor}[1]{\lfloor {#1}\rfloor}
\newcommand{\bfrac}[2]{\left(\frac{#1}{#2}\right)}

\newcommand{\limm}{\lim\limits_{m\rightarrow\infty}}
\newcommand{\lra}{\longrightarrow}
\newcommand{\mtoinf}{m\rightarrow\infty}
\newcommand{\limarm}{\longrightarrow}
\newcommand{\limP}{\stackrel{P}{\longrightarrow}}
\newcommand{\eqD}{\stackrel{\mathcal{D}}{=}}
\newcommand{\abs}[1]{\left| {#1} \right|}

\newcommand{\lilo}[1]{o\left({#1}\right)}

\newcommand{\bigo}[1]{\mathcal{O}\left({#1}\right)}

\newcommand{\op}{o_P(1)}
\newcommand{\opo}[1]{o_P\left({#1}\right)}

\newcommand{\Op}{\mathcal{O}_P(1)}
\newcommand{\Opo}[1]{\mathcal{O}_P\left({#1}\right)}

\newcommand{\ind}[1]{I_{\{{#1}\}}}

\newcommand{\page}{\textnormal{Page}}

\newcommand{\emb}{\overline{\eps}_m}

\newcommand{\kb}{{k^{*}}}
\newcommand{\Nb}{\overline{N}}
\newcommand{\Nd}{N_\delta}
\newcommand{\Nbd}{\Nb_\delta}
\newcommand{\gtilde}{g(m,k)}

\newcommand{\gtildemi}{g(m,i)}

\newcommand{\sbfk}{\sqrt{m}\lr{k/m}^\g}

\newcommand{\VGFN}{\sqrt{N}\lr{k/N}^\g}
\newcommand{\kNg}{\lr{k/N}^\g}
\newcommand{\kNmg}{\lr{k/N}^{-\g}}

\newcommand{\Dm}{\Delta_m}

\newcommand{\vf}{\bfrac{N}{m}^{\g-\oh}} 

\newcommand{\am}{a_m}

\newcommand{\bm}{b_m}

\newcommand{\gkmf}{\left(1+k/m\right)\left(k/(k+m)\right)^\gamma}

\newcommand{\minnik}{\min\limits_{0\leq i\leq k}}
\newcommand{\minniN}{\min\limits_{0\leq i\leq N}}

\newcommand{\maxnik}{\max\limits_{0\leq i\leq k}}
\newcommand{\maxkkb}{\max\limits_{1\leq k< \kb}}
\newcommand{\maxkbkN}{\max\limits_{\kb\leq k\leq N}}
\newcommand{\maxkbkNb}{\max\limits_{\kb\leq j\leq \Nb}}
\newcommand{\maxdelkN}{\max\limits_{\Nd \leq k\leq N}}
\newcommand{\maxdelkNb}{\max\limits_{\Nbd \leq j\leq \Nb}}
\newcommand{\maxkbkdelN}{\max\limits_{\kb \leq k< \Nd}}
\newcommand{\maxkbkdelNb}{\max\limits_{\kb \leq j< \Nbd}}
\newcommand{\maxkN}{\max\limits_{1\leq k\leq N}}

\newcommand{\maxniN}{\max\limits_{0\leq i\leq N}}
\newcommand{\maxnjN}{\max\limits_{0\leq j\leq N}}

\newcommand{\pw}[2]{W_S({#1},{#2})}
\newcommand{\wdr}[1]{W_D(m,{#1})}

\newcommand{\sume}[2]{\sum\limits_{#1 = m+1}^{m+#2}\eps_{#1}}

\newcommand{\sumem}{\sum\limits_{\ell = 1}^{m}\eps_{\ell}}

\newcommand{\Qm}[1]{Q_1(m,#1)}

\newcommand{\Qpil}{S_1(m,k)}

\newcommand{\tp}{\tau^\page_{1,m}}

\newcounter{A-index}
\newcommand{\aind}{\arabic{A-index}}
\newcommand{\aindp}{\stepcounter{A-index}\arabic{A-index}}

\begin{document}
\renewcommand{\thefootnote}{\fnsymbol{footnote}}

\begin{center}
\nin{\Large\textbf{
ASYMPTOTIC DISTRIBUTION OF THE DELAY TIME IN PAGE'S SEQUENTIAL PROCEDURE
}

\renewcommand{\thefootnote}{\arabic{footnote}}}
\end{center}

\begin{center}
\textbf{{Stefan Fremdt$^{\,\mathbf{a}}$
}}
\end{center}

\begin{center}
\medskip
{\normalsize ${}^\text{a}$\,Mathematical Institute, University of Cologne,
\\
Weyertal 86--90, D--50\,931 K\" oln, Germany,\\
e-mail: sfremdt@math.uni-koeln.de\\
Tel.: +49 221 470 2887\\
Fax.: +49 221 470 6073}

\end{center}

\medskip\nin

\begin{abstract}
\noindent 
In this paper the asymptotic distribution of the stopping time in Page's sequential cumulative sum (CUSUM) procedure is presented. Page as well as ordinary cumulative sums are considered as detectors for changes in the mean of observations satisfying a weak invariance principle. The main results on the stopping times derived from these detectors extend a series of results on the asymptotic normality of stopping times of CUSUM-type procedures. In particular the results quantify the superiority of the Page CUSUM procedure to ordinary CUSUM procedures in late change scenarios.
The theoretical results are illustrated by a small simulation study, including a comparison of the performance of ordinary and Page CUSUM detectors.

\vspace{8mm}


\vspace{2mm}

\noindent {\em Abbreviated Title:} Asymptotic distribution of Page's CUSUM

\vspace{2mm}


\end{abstract}

\allowdisplaybreaks

\section{Introduction}\label{sec1}
Monitoring the adequacy of stochastic models is undoubtedly of great importance in many areas of application. A change in the dynamics of the model results in misspecifications and consequently influences conclusions drawn from the model output, e.g., forecasts. Since false conclusions generally imply increasing costs, the speed of detection is crucial for the construction of sequential change-point procedures.\medskip\\
As one of the founding fathers of sequential change-point analysis \citet{1954} introduced the Page CUSUM detector motivated by certain problems in quality control. In this context the properties of detectors were explored mainly employing constant thresholds. In the literature, a comparison of the speed of detection of this type of procedures in many cases is based on their {\it average} detection delay. Even optimality criteria that were introduced for such procedures are referring to the {\it average} run length (ARL) to false alarm. However, procedures designed with respect to this criterion typically stop with (asymptotic) probability one. 
For further reading on optimality criteria for Page's CUSUM we refer to the work of \citet{lorden1971}. In addition the monograph of \citet{BassevilleNikiforov1993} gives an extensive overview of the contributions made since the CUSUM procedure was introduced by \citet{1954}.\medskip\\
\citet{1996} argue that in many contemporary applications, in particular in an economic context, there are no sampling costs under structural stability. They therefore propose an approach, in which the probability of false alarm is controlled. This approach initiated a multitude of contributions to the field of sequential change-point analysis. 
E.g., \citet{2004} propose various sequential tests for the stability of a linear model with independent errors. These tests are based on a stopping rule which stops as soon as a CUSUM detector crosses a given boundary function. In a time series regression model, \citet{F2012-art} proposed ---in a similar fashion--- a stopping rule given by the first-passage time of Page's CUSUM over this very boundary function. Based on these procedures we will consider a location model, i.e., we will investigate changes in the mean of certain times series. The latter model is, as a special case, included in the time series regression model of \citet{F2012-art}.  To compare the performance of the CUSUM procedure of \citet{2004} (which in the sequel will be referred to as {\it ordinary CUSUM}) and the Page CUSUM, we will investigate the limit distributions of the corresponding stopping times for early as well as for late change scenarios. Until now, only few contributions were made regarding the complete asymptotic distribution of the stopping times which obviously provides more information about the behavior of the stopping times than results on the average run length.\medskip\\
Ordinary CUSUM detectors are defined as the partial sum of, e.g., model residuals from the beginning of the monitoring to the present. These procedures have been studied extensively in the literature, cf., e.g., \citet{2004}, \citet{HKS2007} or \citet{AHHK2006}. For the location model results on the asymptotic distribution of the ordinary CUSUM procedure in an early change scenario were given  by \citet{A2003} and \citet{AH2004}. \citet{AHR2007} then also provided an extension to a linear regression model. To prove these results on the asymptotic normality of the ordinary CUSUM detector, strong conditions on the time of change as well as on the magnitude of change were imposed. In this context we also want to mention the work of \citet{huskova2005}, who introduced monitoring procedures based on quadratic forms of weighted cumulative sums, and \citet{HKPS2011}, who showed the asymptotic normality of the corresponding stopping time under assumptions on the time of change similar to those used in \citet{AH2004} and \citet{AHR2007}.\medskip\\ Building on the work of \citet{AH2004}, we will derive the asymptotic distribution of the Page as well as the ordinary CUSUM procedure under weaker conditions on the time and magnitude of the change. Hereby, we will show that the Page procedure is more robust to the location and the size of the change than ordinary CUSUM procedures. The corresponding limit distributions for the Page CUSUM are novel in this context and provide a classification of the behavior of the stopping time depending on the interplay of the magnitude and location of the change.\medskip\\
The paper is organized as follows. In Section \ref{sec3} we will introduce our model setting and required assumptions and formulate our main results. Section \ref{sec4} contains the results of a small simulation study which compares the ordinary and Page CUSUM in various scenarios. Furthermore, it illustrates particularly how the performance of the procedures depends on the time of change. The proofs of our main results from Section \ref{sec3} are postponed to the appendix.
\section{Asymptotic distribution of the stopping times}\label{sec3}
Let $X_i, i = 1,2,\ldots$ follow the location (or ``change-in-the-mean'') model
\begin{equation} X_i = \begin{cases}
          \mu + \eps_i, & i = 1\pp m+\kb - 1,\\
          \mu + \eps_i + \Dm, & i= m+\kb,m+\kb+1,\ldots,
         \end{cases}\label{x_i}
\end{equation} 
where $\mu$ and $\Dm$ are real numbers, $1\leq \kb<\infty$ denotes the unknown time of change and the $\{\eps_i\}_{i=1,2,\ldots}$ are zero-mean random variables. In this setting, the number $m$ is the length of a so-called {\it training period}, in which the model is assumed to be stable. This {\it noncontamination assumption} (cf. \citet{1996}) is used to estimate the model parameters and the asymptotics considered here are (if not stated otherwise) with respect to $m\to\infty$. Furthermore we want to allow for certain dependence structures like autoregressive or GARCH-type dependencies. For this purpose, we assume that the random variables $\eps_i$ satisfy the following weak invariance principles
\begin{align}
 &\left|\sum_{i = 1}^m \eps_i\right| =  \Opo{\sqrt{m}}
 ,\tag{A1}\label{A1}\\
 &\text{There is a sequence of Wiener processes }\{W_m(t):t\geq 0\}_{m\geq 1} \text{ and a positive constant }\sigma\tag{A2}\label{A2}\\
 &\text{such that}\notag\\
 &\sup_{\frac{1}{m}\leq t<\infty}\frac{1}{(mt)^{1/\nu}}\left|\sum_{i=m+1}^{m+mt}\eps_i - \sigma W_m(mt)\right| = \Op \quad 
 \text{ with some }\nu>2. \notag
\end{align}
Examples for sequences of random variables satisfying Assumptions \eqref{A1} and \eqref{A2} are given in \citet{AH2004}. Besides i.i.d.\ sequences these include, e.g., martingale difference sequences and certain stationary mixing sequences. \citet{AHHK2006} showed that the class of augmented GARCH processes, which were introduced by \citet{duan1997}, also satisfies Assumptions \eqref{A1} and \eqref{A2}. This class contains most of the conditionally heteroskedastic time series models applied to describe financial time series. A selection of GARCH models from this class can be found in \citet{ABH2006}.\medskip\\
We are interested in testing the hypothesis
\begin{align}
&H_0: \quad\Dm = 0\notag
\\
\intertext{against the alternatives}
&H_{A,1}:\quad\Dm> 0\qquad \text{and}\qquad H_{A,2}:\quad\Dm\neq 0.\notag
\end{align}
An (asymptotic) $\alpha$-level sequential test of power one (cf. \citet{1996}) is then given via a stopping rule $\tau = \tau_m$ such that
\begin{align*}
 &\limm P(\tau_m < \infty) = \alpha\quad\textnormal{under }H_0 
 \intertext{and}
 &\limm P(\tau_m < \infty) = 1\quad\textnormal{under the alternative.}
\end{align*}

For the location model, \cite{AH2004} defined the CUSUM detector of the (centered) $X_i$
\begin{align*}
\Qm{k} &= \sum_{i=m+1}^{m+k}X_i -\frac{k}{m}\sum_{i = 1}^m X_i, \notag\\
\intertext{and as a corresponding stopping time }
\tau^Q_{1,m} &= \min\{k\geq 1: ~\Qm{k} \geq \hat{\sigma}_mc^Q_{1,\alpha} \gtilde\}.
\end{align*}
Here, $\hat{\sigma}_m$ is a weakly consistent estimator for the constant $\sigma$ from \eqref{A2} (calculated from the data of the training period $1,\ldots,m$),
\begin{equation}
\gtilde = \sqrt{m}\gkmf\quad\text{for }\g\in[0,1/2)\label{g-def} 
\end{equation}
and the critical value $c^Q_{1,\alpha} = c^Q_{1,\alpha}(\gamma)$ is a constant derived from the asymptotic distribution of the detector under the null hypothesis. This asymptotic distribution can be obtained by adapting the proof of Theorem 2.1 in \citet{2004} with respect to Assumptions \eqref{A1} and \eqref{A2}. Under $H_0$, we find
\begin{equation*}
 \lim_{\mtoinf} P\lr{\frac{1}{\hat{\sigma}_m}\Sup_{1\leq k<\infty}\frac{\Qm{k}}{\gtilde}> c^Q_{1,\alpha}} = P\lr{\Sup_{0< t< 1} \frac{W(t)}{t^\g} > c^Q_{1,\alpha}}=\alpha.
\end{equation*}
The parameter $\gamma$ in this construction is a so-called tuning parameter, which allows a practitioner to regulate the sensitivity of the procedure with respect to the time of change. A value of $\gamma$ close to $1/2$ increases detection speed in early change scenarios, while for later changes smaller values lead to a faster detection. For a discussion of this we refer to, e.g., \citet{2004}.\bigskip\\
\citet{AH2004} showed that for the stopping time $\tau^Q_{1,m}$ under the more restrictive {\it local change} assumption $\Dm\to 0$ and for early change alternatives
\begin{align}
 \kb = \bigo{m^\beta}\quad\text{with some }0\leq\beta<\bfrac{\frac{1}{2}-\g}{1-\g}^2, 
 \label{AH-Ass}
\end{align}
one can find (deterministic) sequences $a_m$ and $b_m$ such that $(\tau^Q_{1,m}-a_m)/b_m$ is asymptotically normal.\medskip\\
Our main results, on the one hand, extend the theorem of \citet{AH2004} for the ordinary CUSUM to a wider range of change scenarios, on the other hand, they provide the corresponding limit distribution for the Page CUSUM. Not only do we weaken the assumptions to allow for a larger class of change-size scenarios, but the range for the value $\beta$ in \eqref{AH-Ass} is extended to the upper limit 1 (see Assumption \eqref{A3} below). 
These results permit a comparison of the two procedures on a theoretical basis. From this the superiority of the Page versus the ordinary CUSUM in late change scenarios can be seen, while the similarity of these procedures in early change scenarios is implied. \bigskip\\
The Page CUSUM detector for the one-sided alternative $H_{A,1}$ is given via
\[\Qpil = \Qm{k} - \min\limits_{0\leq i\leq k} \Qm{i}\qquad\textnormal{(with }\Qm{0} = 0)\]
and the corresponding stopping time with $g$ from \eqref{g-def} as
\[\tp = \min\{k\geq 1: ~\Qpil{} \geq \hat{\sigma}_m c^{\page}_{1,\alpha} \gtilde\}.\]
Here, according to \citet{F2012-art}, the critical value $c^{\page}_{1,\alpha}=c^{\page}_{1,\alpha}(\g)$ for a given confidence level $\alpha \in (0,1)$ can be chosen such that under $H_0$
\begin{align*}
 &\lim_{\mtoinf} P\lr{\frac{1}{\hat{\sigma}_m}\Sup_{1\leq k<\infty}\frac{\Qpil}{\gtilde}> c^{\page}_{1,\alpha}} \\
 = &P\lr{\Sup_{0< t< 1} \ofrac{t^\g}\left[W(t)-\Inf_{0\leq s\leq t}\lr{\frac{1-t}{1-s}W(s)}\right] > c^{\page}_{1,\alpha}}\notag\\
 =&\alpha.\notag
\end{align*}
A table containing simulated versions of the critical values $c^{\page}_{1,\alpha}(\g)$ for selected values of $\g$ and $\alpha$ can be found in \citet{F2012-art}.\medskip\\
It can easily be seen that the stopping times $\tau^Q_{1,m}$ and $\tp$ are suitable for alternative $H_{A,1}$. The (two-sided) detectors and stopping rules for $H_{A,2}$ are given via
\begin{align*}
 &S_2(m,k) = \maxnik |\Qm{k} - \Qm{i}|,\quad\tau^{\page}_{2,m} = \min\{k\geq 1: ~S_2(m,k) \geq \hat{\sigma}_m c^{\page}_{2,\alpha} \gtilde\}
 \intertext{and}
 &Q_2(m,k) = |\Qm{k}|,\quad\tau^Q_{2,m} = \min\{k\geq 1: ~Q_2(m,k) \geq \hat{\sigma}_mc^Q_{2,\alpha} \gtilde\}.
\end{align*}
For a proof of consistency and the derivation of the limit distributions under $H_0$ we refer to \citet{F2012-art} and \citet{2004}.\medskip\\
The asymptotic distribution of the stopping times $\tp$ and $\tau^{\page}_{2,m}$ will depend strongly on the interplay of $\Dm$ and $\kb$. To specify this, we need the following (technical) assumptions:
\begin{align}
&\quad \textnormal{there exists a }\theta>0\text{ such that }\kb = \lfloor\theta m^\beta\rfloor\text{ with }0\leq\beta<1\tag{A3}\label{A3}\\
&\quad\sqrt{m}|\Dm|\limarm \infty,\tag{A4} \label{A4}\\
&\quad\Dm = \bigo{1}.\tag{A5}\label{A5}
\end{align}
Assumption \eqref{A5} only requires $\Dm$ to be bounded and therefore includes {\it local} as well as {\it fixed} alternatives. From Assumption \eqref{A3}, we have a given order of the change-point $\kb$ in terms of $m$ depending on the exponent $\beta$. As we will see later on, the asymptotic distribution of the stopping times depends crucially on the decay of $\Dm$, which is implicitly allowed by Assumptions \eqref{A4} and \eqref{A5}. This dependence can be expressed in terms of the asymptotic behavior of the quantities $|\Dm| m^{\g-1/2}\kb^{1-\g}$. In view of Assumption \eqref{A3} we distinguish the following three cases:
\begin{align}
& m^{\beta(1-\g)-\oh+\g}|\Dm|\limarm 0, \tag{I}\label{I}\\
& m^{\beta(1-\g)-\oh+\g}|\Dm|\limarm \tilde{C}_1\in (0,\infty),
\tag{II}\label{II}\\
& m^{\beta(1-\g)-\oh+\g}|\Dm|\limarm \infty. \tag{III}\label{III}
\end{align}

\begin{re}\label{re1}
\begin{enumerate}[a)]
 \item  We note that under Assumptions \eqref{A4} and \eqref{A5}) we have \eqref{I} for
 \begin{align}
  0\leq \beta < \frac{\frac 12 - \g}{1-\g} \tag{Ia}\label{Ia}.
 \end{align}

 \item  Assumption \eqref{A3} implies that under \eqref{II} we have
 \begeq
 {|\Dm| {m^{\g-\oh}}\kb^{1-\g}}\limarm \theta^{1-\g}\tilde{C}_1 = C_1\in(0,\infty).\label{re1-eq2}
 \eneq
\end{enumerate}
\end{re}
The limit distribution in our main results will depend on the given case \eqref{I}, \eqref{II} or \eqref{III}. In order to define this limit distribution, we first introduce for all real $x$
\begeqO
\overline{\Psi}(x) = \begin{cases}
 \Phi(x), &\text{under }\eqref{I},\\
 P\lr{\Sup_{d_1 < t<1}W(t)\leq x},&\text{under} \eqref{II},\\
 P\lr{\Sup_{0<t<1}W(t)\leq x} = \begin{cases}
                                 0,&x<0,\\
                                 2\Phi(x) - 1,& x\geq 0,
                                \end{cases} 

,&\text{under }\eqref{III}.
\end{cases}
\eneqO
Here $\Phi(x)$ denotes the standard normal distribution function and under \eqref{II} we denote by $d_1 = d_1(c)$ the unique solution of 
\begin{align}
d_1 = 1 - \frac{c\sigma}{C_1}d_1^{1-\g}.\label{d1} 
\end{align}
\begin{Th}\label{AD}
Let $\{X_n\}_{n\geq 1}$ be a sequence of random variables according to \eqref{x_i} such that \eqref{A1} -- \eqref{A5} are satisfied, and let $\g\in[0,1/2)$.
\begin{enumerate}[(a)]
 \item Then, for all real $x$ under $H_{A,1}$ and with $\Psi(x)= 1-\overline{\Psi}(-x)$,
 \begin{equation*}
\limm P\left(\frac{\tp - \am(c^{\page}_{1,\alpha})}{\bm(c^{\page}_{1,\alpha})}\leq x\right) =\Psi(x),
\end{equation*}
where $\am(c)$ is the unique solution of 
\begin{align*}
\am(c) &= \lr{\frac{\sigma cm^{\oh-\g}}{|\Dm|}+\frac{\kb}{(\am(c))^\g}}^{1/(1-\g)}\mspace{-30mu}\\
\intertext{and }\bm(c) &= \sigma\sqrt{\am(c)}|\Delta_m|^{-1}\lr{1-\g\lr{1-\frac{\kb}{\am(c)}}}^{-1}\mspace{-23mu}.
\end{align*}
\item Additionally it holds that, for all real $x$ under $H_{A,1}$,
\begin{equation*}
\limm P\left(\frac{\tau^Q_{1,m} - \am(c^Q_{1,\alpha})}{\bm(c^Q_{1,\alpha})}\leq x\right) = \Phi(x).
\end{equation*}
\end{enumerate}
\end{Th}
\begin{Th}\label{AD2}
Let $\{X_n\}_{n\geq 1}$ be a sequence of random variables according to \eqref{x_i} such that \eqref{A1} -- \eqref{A5} are satisfied and let $\g\in[0,1/2)$.
Then, for all real $x$ under $H_{A,2}$, the limit results of Theorem \ref{AD} are retained if $\tp, c^{\page}_{1,\alpha}, \tau^Q_{1,m}$ and $c^Q_{1,\alpha}$ are replaced with the respective objects $\tau^{\page}_{2,m}, c^{\page}_{2,\alpha}, \tau^Q_{2,m}$ and $c^Q_{2,\alpha}$.
\end{Th}
Theorems \ref{AD} and \ref{AD2} show that in early change scenarios like under case \eqref{Ia}, the behavior of the ordinary and Page CUSUM is comparable. This implies that the same limit distribution is obtained using normalizing sequences which differ only by the critical values. As, clearly, $c^{\page}_{1,\alpha}>c^Q_{1,\alpha}$ this suggests a slightly superior behavior of the ordinary CUSUM.\medskip\\
Under late change scenarios, however, the difference in the limit distribution shows how the detection delay improves for the Page CUSUM, since the Page limit distribution is stochastically smaller than a standard normal one (cf.\ \eqref{I} compared to \eqref{II} and \eqref{III}). This underlines the detection properties of the Page CUSUM which by construction is intended to show a higher stability in particular under later changes. Furthermore, the construction of the normalizing sequences $a_m(c)$ and $b_m(c)$ identifies the driving factors for the detection delay and therefore adds to the understanding of the dynamics of the different procedures.\medskip\\
In the next section we will find that the large sample results from above can be confirmed empirically. In addition they can help to explain effects observed in the small sample behavior of the considered procedures.
\section{A small simulation study}\label{sec4}
\begin{figure}
 \centering
 \includegraphics[page=1, scale = 0.8]{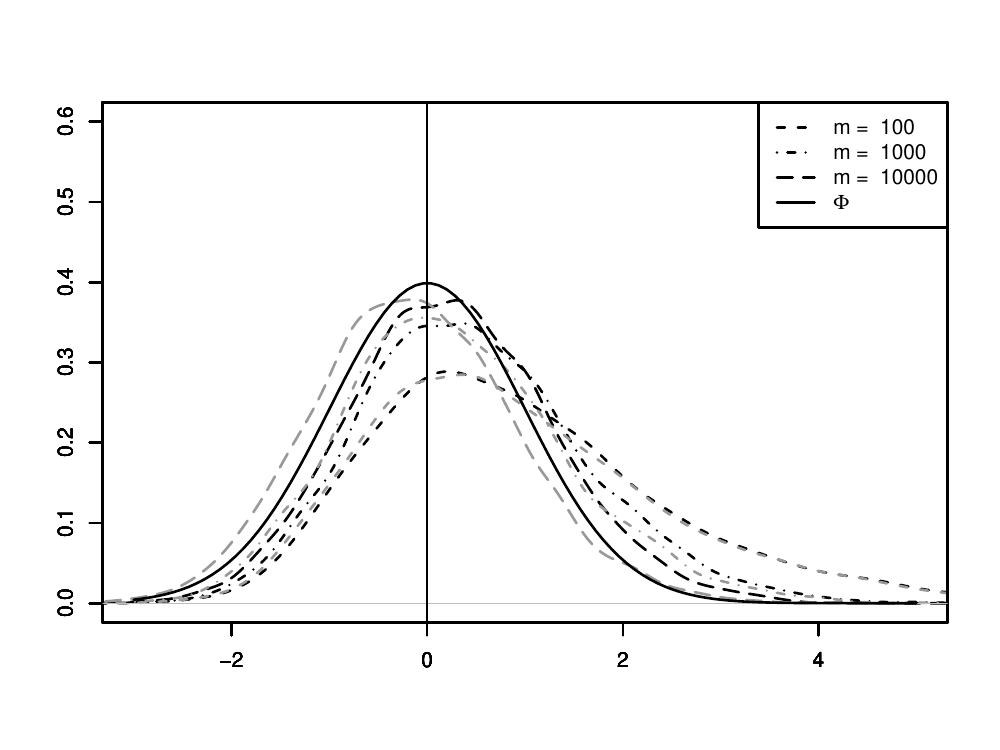}
 \includegraphics[page=2, scale = 0.8]{plot_densities_Fig-1.pdf}
 \includegraphics[page=3, scale = 0.8]{plot_densities_Fig-1.pdf}
 \includegraphics[page=4, scale = 0.8]{plot_densities_Fig-1.pdf}
 \includegraphics[page=5, scale = 0.8]{plot_densities_Fig-1.pdf}
 \includegraphics[page=6, scale = 0.8]{plot_densities_Fig-1.pdf}
 \caption{Estimated density plots for $\kb = \lfloor \theta m^0\rfloor$ with $\theta = 1$ for $\nu^{\page}_{1,m}$ (left column) and $\nu^Q_{1,m}$ (right column) in black lines. In both columns the estimated density of $\tilde{\nu}_m$ can be found in gray lines. The rows from top to bottom correspond to the tuning parameter  $\g = 0.00,0.25$ and $0.45$.}
\label{fig:1-1}
\end{figure}

\begin{figure}
 \centering
 \includegraphics[page=7, scale = 0.8]{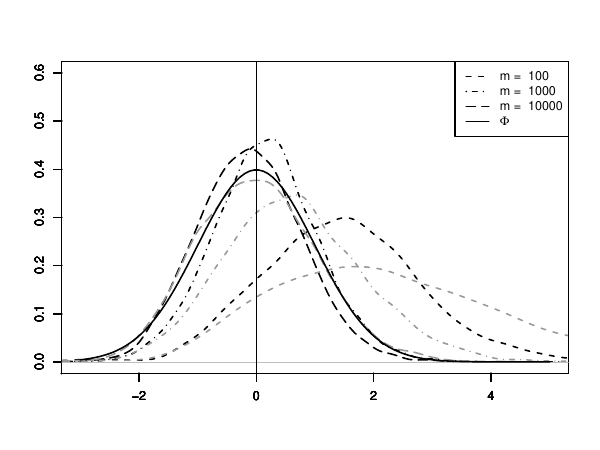}
 \includegraphics[page=8, scale = 0.8]{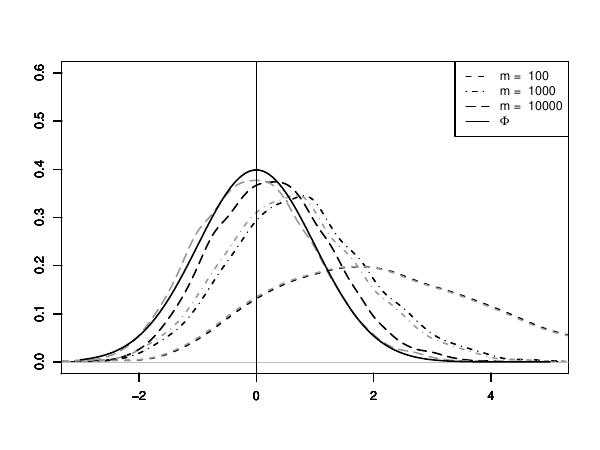}
 \includegraphics[page=9, scale = 0.8]{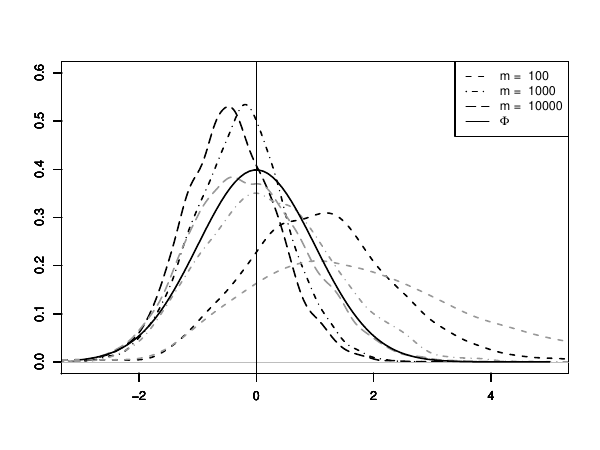}
 \includegraphics[page=10, scale = 0.8]{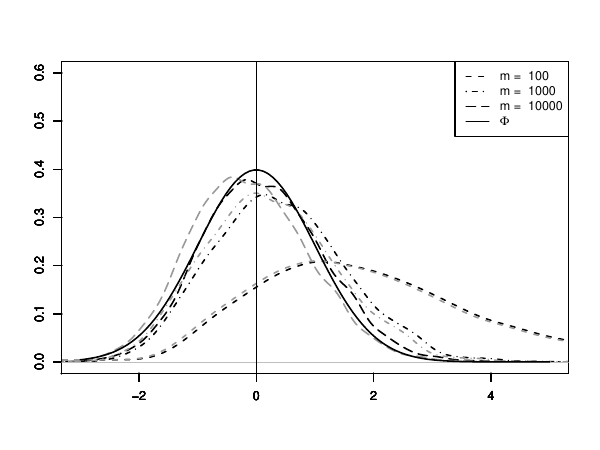}
 \includegraphics[page=11, scale = 0.8]{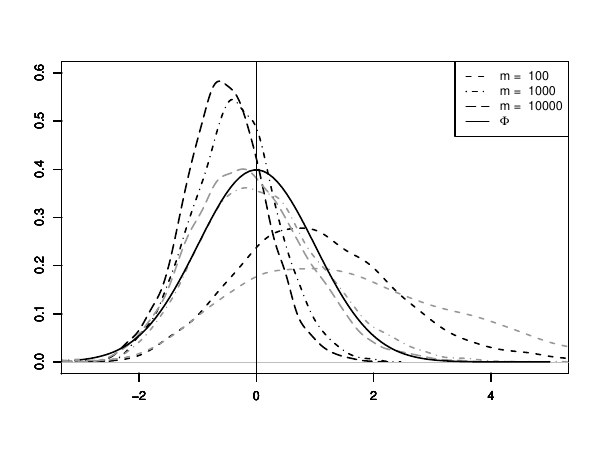}
 \includegraphics[page=12, scale = 0.8]{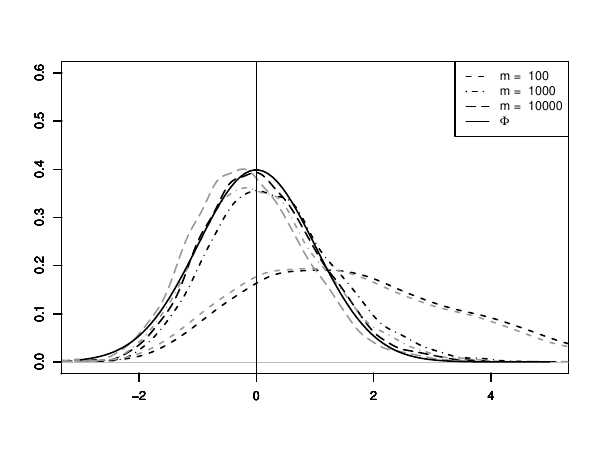}
 \caption{Estimated density plots for $\kb = \lfloor \theta m^0\rfloor$ with $\theta = 100$ for $\nu^{\page}_{1,m}$ (left column) and $\nu^Q_{1,m}$ (right column) in black lines. In both columns the estimated density of $\tilde{\nu}_m$ can be found in gray lines. The rows from top to bottom correspond to the tuning parameter  $\g = 0.00,0.25$ and $0.45$.}
\label{fig:1-1a}
\end{figure}

\begin{figure}
 \centering
 \includegraphics[page=1, scale = 0.8]{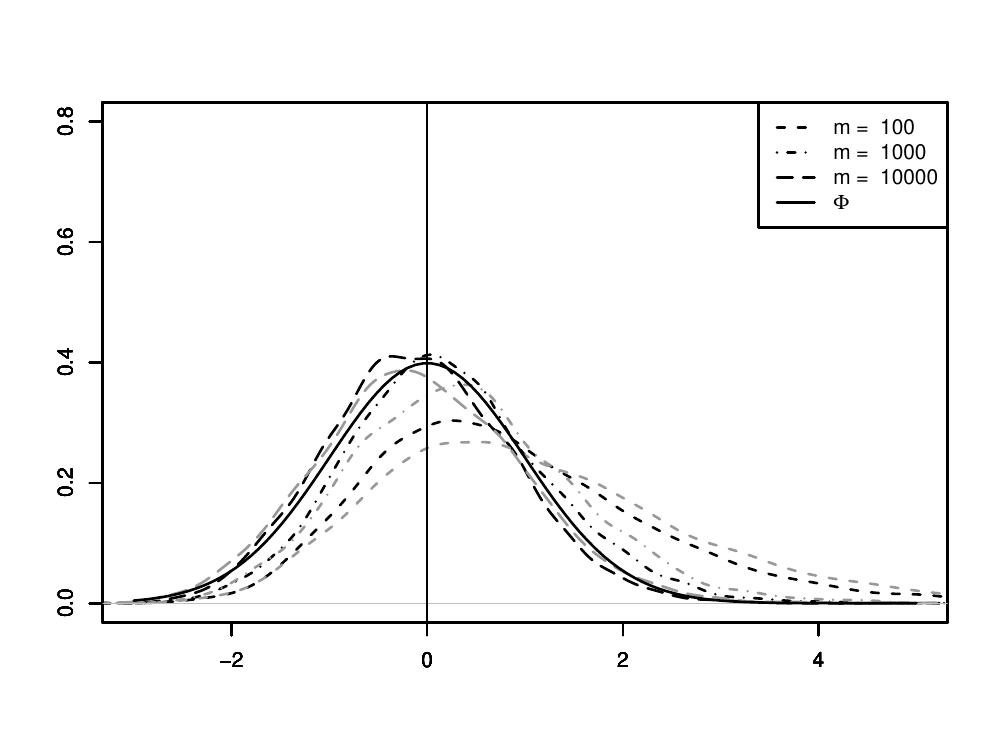}
 \includegraphics[page=2, scale = 0.8]{plot_densities_Fig-2.pdf}
 \includegraphics[page=3, scale = 0.8]{plot_densities_Fig-2.pdf}
 \includegraphics[page=4, scale = 0.8]{plot_densities_Fig-2.pdf}
 \includegraphics[page=5, scale = 0.8]{plot_densities_Fig-2.pdf}
 \includegraphics[page=6, scale = 0.8]{plot_densities_Fig-2.pdf}
  \caption{Estimated density plots for $\kb = \lfloor m^{0.45}\rfloor$ for $\nu^{\page}_{1,m}$ (left column) and $\nu^Q_{1,m}$ (right column) in black lines. In both columns the estimated density of $\tilde{\nu}_m$ can be found in gray lines. The rows from top to bottom correspond to the tuning parameter  $\g = 0.00,0.25$ and $0.45$.}
%
\label{fig:1-2}
\end{figure}

\begin{figure}
 \centering
 \includegraphics[page=1, scale = 0.8]{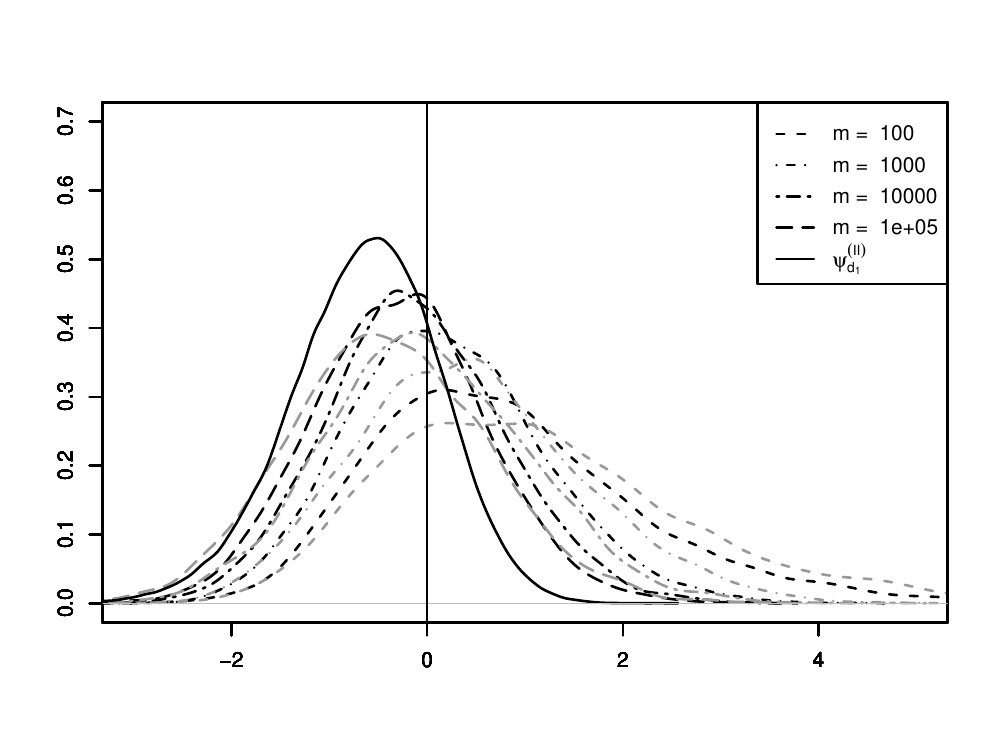}
 \includegraphics[page=2, scale = 0.8]{plot_densities_Fig-3.pdf}
 \includegraphics[page=3, scale = 0.8]{plot_densities_Fig-3.pdf}
 \includegraphics[page=4, scale = 0.8]{plot_densities_Fig-3.pdf}
 \includegraphics[page=5, scale = 0.8]{plot_densities_Fig-3.pdf}
 \includegraphics[page=6, scale = 0.8]{plot_densities_Fig-3.pdf}
 \caption{Estimated density plots for $\nu^{\page}_{1,m}$ (left column) and $\nu^Q_{1,m}$ (right column) in black lines. In both columns the estimated density of $\tilde{\nu}_m$ can be found in gray lines. The rows from top to bottom correspond to the tuning parameter  $\g = 0.00,0.25$ and $0.45$. In each row the change-point $\kb$ was set to $\kb = \lfloor m^\beta\rfloor$ such that $\beta = (1/2-\g)/(1-\g)$. The limit distribution in the left column is therefore determined by case \eqref{II}.}
%
\label{fig:2}
\end{figure}

\begin{figure}
 \centering
 \includegraphics[page=1, scale = 0.8]{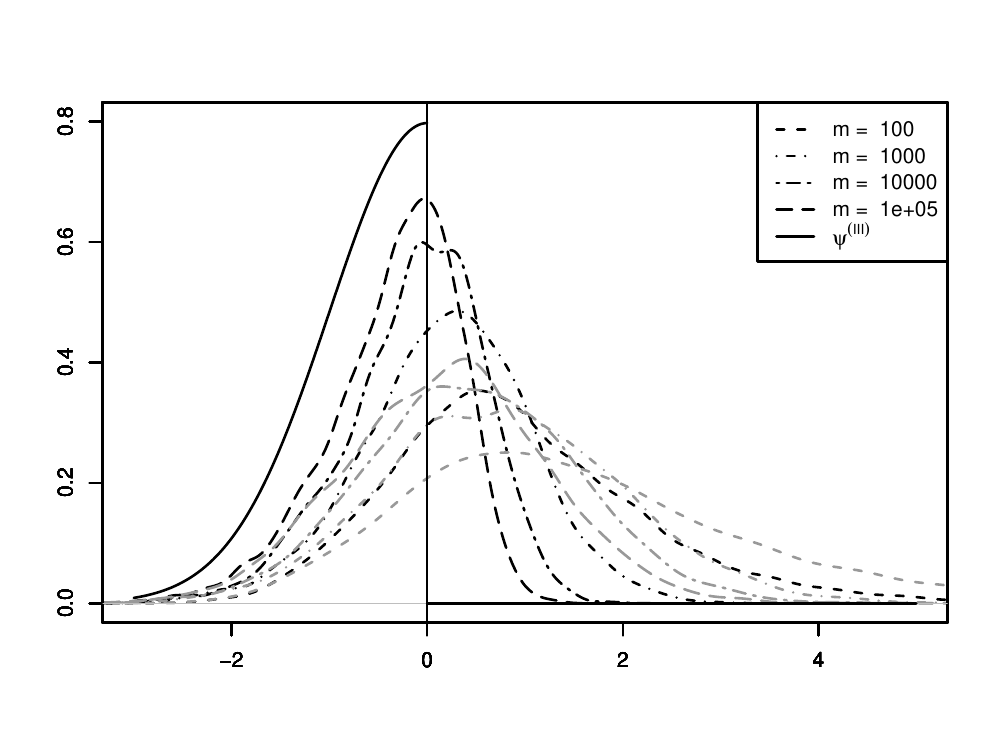}
 \includegraphics[page=2, scale = 0.8]{plot_densities_Fig-4.pdf}
 \includegraphics[page=3, scale = 0.8]{plot_densities_Fig-4.pdf}
 \includegraphics[page=4, scale = 0.8]{plot_densities_Fig-4.pdf}
 \includegraphics[page=5, scale = 0.8]{plot_densities_Fig-4.pdf}
 \includegraphics[page=6, scale = 0.8]{plot_densities_Fig-4.pdf}
\caption{Estimated density plots for $\kb = \lfloor m^{0.75}\rfloor$ for $\nu^{\page}_{1,m}$ (left column) and $\nu^Q_{1,m}$ (right column) in black lines. In both columns the estimated density of $\tilde{\nu}_m$ can be found in gray lines. The rows from top to bottom correspond to the tuning parameter  $\g = 0.00,0.25$ and $0.45$.}

\label{fig:4}
\end{figure}

In this section, we present the outcome of a small simulation study to illustrate the theoretical findings from Section \ref{sec3} and compare ordinary and Page CUSUM in this framework. The simulations were carried out for various types of sequences $\{\eps_i\}_{i=1,2,\ldots}$, mainly leading to similar findings. Since the statements of Theorems \ref{AD} and \ref{AD2} yield the behavior of ordinary and Page CUSUM in different scenarios, we will focus here on these scenarios and the transition between them. We will therefore only provide one examplary setting. A more extensive empirical analysis of the limit behavior with respect to the influence of the different parameters and the dependence structure of the series $\{\eps_i\}_{i=1,2,\ldots}$ would be desirable. Yet this exceeds by far the scope of this paper. However, it should be mentioned that, in particular in small samples, the quality of estimation of the parameter $\sigma$ is crucial to the behavior of the procedures. Inaccurate estimation may lead to increasing sizes under the null hypothesis, which bias the estimated densities of the stopping times as well. \medskip\\
We will focus here on results for $\mu = 0$, using a GARCH(1,1) sequence 
\begin{align*}
&\eps_i = \sigma_i z_i,\quad\sigma_i^2 = \overline{\omega} + \overline{\alpha}\eps_{i-1}^2 + \overline{\beta}\sigma_{i-1}^2,
\intertext{where $\{z_i\}_{i = 1,2,\ldots}$ are i.i.d. standard normally distributed and the parameters were specified as}
 &\overline{\omega} = 0.5,~\overline{\alpha} = 0.2~\text{and}~\overline{\beta} = 0.3, 
\end{align*}
which implies (unconditional) unit variance. Due to the uncorrelated error terms, the ordinary least squares estimator for the sample variance is a weakly consistent estimator for the parameter $\sigma$. All simulations were carried out using 5000 replications.\medskip\\
Since the limit behavior of the stopping times depends strongly on the interplay of the model parameters, a rather simple simulation design was chosen, to increase the traceability and make the outcome easier to interpret. For all presented results a fixed change alternative with $\Dm = 1$ was considered, which implies that the behavior of $m^{\beta(1-\g)-1/2+\g}\Dm$, and hereby the determination of the corresponding case \eqref{I} -- \eqref{III} depends only on the exponent $\eta = \eta(\g,\beta) = \beta(1-\g)-1/2+\g$. Hence, the cases correspond to $\eta\lesseqqgtr 0$. In all presented figures, the respective density of the limit distribution is plotted as a solid line and denoted by $\Phi(=\Psi^{\text{(I)}}), \Psi^{\text{(II)}}_{d_1}$ and $\Psi^{\text{(III)}}$. Since the one and two-sided procedures mainly differ in the critical values, we only present the results for the one-sided procedures.\medskip\\
Figures \ref{fig:1-1}--\ref{fig:4} each show (in black lines) estimated density plots of the normalized stopping time 
\begin{align*}
 \nu^{\page}_{1,m} &= (\tp - a_m(c^{\page}_{1,\alpha}))/b_m(c^{\page}_{1,\alpha})
\end{align*}
in the left column and
\begin{align*}
 \nu^Q_{1,m} &= (\tau^Q_{1,m} - a_m(c^Q_{1,\alpha}))/b_m(c^Q_{1,\alpha})
\end{align*}
in the right column.
%
These illustrate the limit results of Theorem \ref{AD} (a) and (b), respectively. For a direct comparison of ordinary and Page CUSUM the normalized stopping time 
\begin{equation*}
\tilde{\nu}_m = (\tau^Q_{1,m} - a_m(c^{\page}_{1,\alpha}))/b_m(c^{\page}_{1,\alpha}) 
\end{equation*}
can be used as a benchmark. In both columns of Figures \ref{fig:1-1}--\ref{fig:4} the corresponding density estimates for $\tilde{\nu}_m$ are provided in gray lines. The rows in Figures \ref{fig:1-1}--\ref{fig:4} correspond to the choice of the tuning parameter $\g$, where from top to bottom, $\g = 0.00,0.25$ and $0.45$. We will first interpret the findings with respect to the limit results themselves and then conclude the analysis with a comparison of $\tp$ and $\tau^Q_{1,m}$ in the different scenarios.\medskip\\
Since the limit distribution of $\nu^Q_{1,m}$ is standard normal under the assumptions of Theorem \ref{AD}, it is not surprising that we will find the convergence to be relatively robust with respect to the given scenario. This, however, does not hold for the convergence of $\tp$, where the transition from case \eqref{I} to case \eqref{III} strongly influences the convergence even for large sample sizes. We will therefore restrict the following discussion to the behavior of the Page CUSUM.\medskip\\
Figure \ref{fig:1-1} shows the estimated density plots for a fixed change-point, i.e., $\beta=0$ in \eqref{A3}, choosing $\kb = \theta = 1$. Figure \ref{fig:1-1a} provides the related results for $\kb = 100$. These belong to case \eqref{I} for all values of $\g$ and therefore have the standard normal distribution as a limit (for $\nu^{\page}_{1,m}$ and $\nu^Q_{1,m}$). For $\kb = 1$ a fast and clear convergence to the standard normal can be found for $\g = 0.00$ and $\g = 0.25$. For $\g = 0.45$, a deviation is visible due to a slower convergence to the standard normal. This deviation can be explained in terms of the transition from case \eqref{I} to case \eqref{II} (and \eqref{III}) in $\eta = 0$ and thus for $\tilde{\beta} = 1/11$. Consider, e.g., the alternative scenario $\kb = \lfloor m^{0.1}\rfloor$. For $m = 100$ and $m = 1000$ we then find $\kb = 1$, for $m = 10,000$ we have $\kb = 2$. The two scenarios are therefore rather indistinguishable for the given sample sizes. Yet they have different limit distributions. Consequently, a fast convergence to the standard normal can hardly be expected. The influence of the parameter $\theta$ from Assumption \eqref{A3}, which leads to a bias in the limiting behavior, is shown in Figure \ref{fig:1-1a}. While for $\g = 0.00$ the convergence to the standard normal distribution can be seen nicely, for the larger values of $\g$ it is not obvious. The explanation for this is again the influence of $\g$ on $\eta$, e.g., we have for $m = 10,000$ that $k^*_1 = 100m^0$ and $k^*_2 = m^{0.5}$ are two possible alternatives with different limit distributions for larger values of $\g$. The parameter $\theta$ was used in earlier works to justify the assumption of an early change scenario. The results from Figure \ref{fig:1-1a} show, however, that this argument has to be handled with caution.\medskip\\
Figure \ref{fig:1-2} shows the density plots for $\kb = \lfloor m^{0.45}\rfloor$, which implies $\eta<0$ for $\g = 0.00$ and $\eta>0$ for $\g = 0.25$ and $\g = 0.45$. For $\g = 0.00$ the convergence to the standard normal distribution is again obvious. For $\g = 0.25$ and $\g = 0.45$, the convergence away from the standard normal distribution towards $\Psi^{\text{(III)}}$ is also clearly visible for $\nu^{\page}_{1,m}$. Figure \ref{fig:2} shows results under case \eqref{II}, which corresponds to $\eta = 0$ and $\beta$ taking values $\beta = 1/2, 1/3$ and $1/11$, for $\g = 0.00,0.25,0.45$. The limiting densities in the left column show the dependence on $d_1$ and therefore implicitly on $\g$. The model setting implies $C_1 = 1$, consequently $d_1$ takes the following values:
\begin{center}
\begin{tabular}{c|c|c|c}
$\g$ & 0.00 & 0.25 & 0.45\\\hline
$d_1(\g)$ &  0.3714 & 0.1887 & 0.1051
 \end{tabular}
 \end{center}
Finally, in Figure \ref{fig:4} case \eqref{III} is considered. The change-point is set to $\kb = \lfloor m^{0.75}\rfloor$ and we find that in accordance with the theoretical results only little influence of $\g$ can be seen. The convergence to the limit distribution $\Psi^{\text{(III)}}$ is again obvious in the left column.

\paragraph{Comparison of $\tp$ and $\tau^Q_{1,m}$.}
Now we want to compare the stopping rules $\tp$ and $\tau^Q_{1,m}$ explicitly. Aside from the large sample behavior provided in Theorem \ref{AD}, the simulations yield further information on the small sample behavior. For this, the discussion of the convergence from above helps to explain the findings with respect to this comparison.\medskip\\
Since the normalizations in Theorem \ref{AD} allow for a direct comparison of the relative values, we will not discuss the absolute values of the delay times and, in particular, the influence of the parameter $\g$ on these. Table \ref{am-bm-table} shows the values of the normalizing sequences $a_m(c^{\page}_{1,\alpha})$, $b_m(c^{\page}_{1,\alpha})$ and $a_m(c^Q_{1,\alpha})$, $b_m(c^Q_{1,\alpha})$ for $\alpha = 0.1$ in the considered scenarios. These confirm the findings of \citet{F2012-art} and \citet{2004}, which is why we refer to the latter works for a further discussion of this topic.\medskip\\
To compare the stopping times $\tp$ and $\tau^Q_{1,m}$ directly, we consider the same normalization for both, i.e., we compare $\nu^{\page}_{1,m}$ and $\tilde{\nu}_m$. Thus, the left columns of Figures \ref{fig:1-1}--\ref{fig:4} provide the basis for this comparison.\medskip\\
As to be expected, Figure \ref{fig:1-1} shows that for a change-point at the beginning of the monitoring, the ordinary CUSUM detects slightly faster than the Page CUSUM. This is mainly due to the difference of the critical values. In Figure \ref{fig:1-1a} it can be seen how the parameter $\theta$ influences the detection of the two procedures. While both have the same limit distribution, the Page CUSUM shows a faster detection in small as well as large samples. Additionally, for $\gamma > 0$ the shape of the density estimates of $\nu^{\page}_{1,m}$ yields the sensitivity of the procedure to the transition between the different cases.\medskip\\
Figure \ref{fig:1-2} particularly highlights the dependence of the limit distribution on the parameter $\g$. While for $\g = 0$ only a slightly faster detection of the Page CUSUM can be observed, for larger values of $\g$ the effect is more evident, due to the difference in the limit distribution.\medskip\\
The decreasing difference of the two procedures with increasing $\g$ in Figure \ref{fig:2} is owed to the change-point specification. For $\g = 0$ the change-point $\kb$ (for $m = 100,1000,10000$) takes the values $\kb = 10,31,100$, for $\g = 0.25$ we have $\kb = 4,9,21$ and for $\g = 0.45$ finally $\kb = 1,1,2$. The observed behavior is therefore not surprising.\medskip\\
Finally, Figure \ref{fig:4} clearly shows how the Page CUSUM improves compared to the ordinary CUSUM, the later a change occurs. Even for small training samples the density estimates of the Page CUSUM dominate those of the ordinary CUSUM and the effect is reinforced with increasing sample size.\medskip\\
To summarize, we find that the simulations in general confirm the theoretical results. They even explain why in certain (biased) early change scenarios, despite the same standard normal limit for both procedures, the Page CUSUM outperforms the ordinary CUSUM. Furthermore, the similarity of the Page and ordinary CUSUM in early change scenarios, on the one hand, and the superiority of the Page CUSUM in late change scenarios, on the other hand, underline why in general (i.e., without additional assumptions on the change-point) the usage of the Page CUSUM procedure can be recommended.
\begin{table}[hptb]
\begin{center}
\begin{tabular}{cccccccccc} 
&&&\multicolumn{3}{c}{$\tp$}&&\multicolumn{3}{c}{$\tau^Q_{1,m}$}\\
\cmidrule(lr){3-6}\cmidrule(lr){7-10}
$\kb$&$\g$&$m$&100&1000&10000&$m$&100&1000&10000\\
\cmidrule(lr){1-6}\cmidrule(lr){7-10}
1&0.00&$a_m$&~~17.92&~~54.52&~170.24&$a_m$&~~17.45&~~53.01&~165.49\\ 
&&$b_m$&~~~4.23&~~~7.38&~~13.05&$b_m$&~~~4.18&~~~7.28&~~12.86\\
\cmidrule(lr){2-6}\cmidrule(lr){7-10}
&0.25&$a_m$&~~12.23&~~24.84&~~52.00&$a_m$&~~11.55&~~23.39&~~48.86\\ 
&&$b_m$&~~~4.54&~~~6.56&~~~9.55&$b_m$&~~~4.41&~~~6.36&~~~9.26\\ 
\cmidrule(lr){2-6}\cmidrule(lr){7-10}
&0.45&$a_m$&~~~9.54&~~11.37&~~13.63&$a_m$&~~~8.57&~~10.18&~~12.15\\ 
&&$b_m$&~~~5.17&~~~5.72&~~~6.33&$b_m$&~~~4.86&~~~5.37&~~~5.94\\ 
\cmidrule(lr){1-6}\cmidrule(lr){7-10}
100&0.00&$a_m$&~116.92&~153.52&~269.24&$a_m$&~116.45&~152.01&~264.49\\
&&$b_m$&~~10.81&~~12.39&~~16.41&$b_m$&~~10.79&~~12.33&~~16.26\\
\cmidrule(lr){2-6}\cmidrule(lr){7-10}
&0.25&$a_m$&~119.87&~136.51&~168.42&$a_m$&~118.90&~134.68&~164.87\\
&&$b_m$&~~11.42&~~12.52&~~14.44&$b_m$&~~11.36&~~12.40&~~14.24\\
\cmidrule(lr){2-6}\cmidrule(lr){7-10}
&0.45&$a_m$&~127.43&~131.18&~135.49&$a_m$&~125.31&~128.75&~132.70\\
&&$b_m$&~~12.50&~~12.82&~~13.20&$b_m$&~~12.31&~~12.61&~~12.96\\
\cmidrule(lr){1-6}\cmidrule(lr){7-10}
$\lfloor m^{0.45}\rfloor$&0.00&$a_m$&~~23.92&~~75.52&~232.24&$a_m$&~~23.45&~~74.01&~227.49\\
&&$b_m$&~~~4.89&~~~8.69&~~15.24&$b_m$&~~~4.84&~~~8.60&~~15.08\\
\cmidrule(lr){2-6}\cmidrule(lr){7-10}
&0.25&$a_m$&~~19.64&~~50.47&~126.72&$a_m$&~~18.94&~~48.92&~123.32\\
&&$b_m$&~~~5.28&~~~8.27&~~12.88&$b_m$&~~~5.17&~~~8.11&~~12.65\\
\cmidrule(lr){2-6}\cmidrule(lr){7-10}
&0.45&$a_m$&~~18.51&~~40.34&~~92.96&$a_m$&~~17.41&~~38.75&~~90.53\\
&&$b_m$&~~~5.97&~~~7.98&~~11.28&$b_m$&~~~5.71&~~~7.73&~~11.02\\
\cmidrule(lr){1-6}\cmidrule(lr){7-10}
$\lfloor m^{0.5}\rfloor$&0.00&$a_m$&~~26.92&~~84.52&~269.24&$a_m$&~~26.45&~~83.01&~264.49\\ &&$b_m$&~~~5.19&~~~9.19&~~16.41&$b_m$&~~~5.14&~~~9.11&~~16.26\\
\cmidrule(lr){1-6}\cmidrule(lr){7-10}
$\lfloor m^{1/3}\rfloor$&0.25&$a_m$&~~16.01&~~34.97&~~77.32&$a_m$&~~15.33&~~33.49&~~74.11\\
&&$b_m$&~~~4.93&~~~7.26&~~10.75&$b_m$&~~~4.80&~~~7.08&~~10.49\\
\cmidrule(lr){1-6}\cmidrule(lr){7-10}
$\lfloor m^{1/11}\rfloor$&0.45&$a_m$&~~~9.54&~~11.37&~~15.30&$a_m$&~~~8.57&~~10.18&~~13.81\\
&&$b_m$&~~~5.17&~~~5.72&~~~6.43&$b_m$&~~~4.86&~~~5.37&~~~6.04\\
\cmidrule(lr){1-6}\cmidrule(lr){7-10}
$\lfloor m^{0.75}\rfloor$&0.00&$a_m$&~~47.92&~230.52&1169.24&$a_m$&~~47.45&~229.01&1164.49\\ &&$b_m$&~~~6.92&~~15.18&~~34.19&$b_m$&~~~6.89&~~15.13&~~34.12\\
\cmidrule(lr){2-6}\cmidrule(lr){7-10}
&0.25&$a_m$&~~46.70&~218.04&1109.61&$a_m$&~~45.90&~216.03&1104.35\\
&&$b_m$&~~~7.46&~~15.50&~~34.15&$b_m$&~~~7.37&~~15.39&~~34.04\\
\cmidrule(lr){2-6}\cmidrule(lr){7-10}
&0.45&$a_m$&~~48.81&~216.02&1090.74&$a_m$&~~47.33&~213.06&1084.14\\
&&$b_m$&~~~8.36&~~16.00&~~34.31&$b_m$&~~~8.14&~~15.80&~~34.12
\end{tabular}
\caption{
Values of the normalizing sequences $a_m(c^{\page}_{1,\alpha})$, $b_m(c^{\page}_{1,\alpha})$ (left) and $a_m(c^Q_{1,\alpha})$, $b_m(c^Q_{1,\alpha})$ (right) for $\alpha = 0.1$ in the scenarios considered in Figures \ref{fig:1-1}--\ref{fig:4}.
}
\label{am-bm-table}
\end{center}
\end{table}
\appendix
\section{Proofs of Theorems \ref{AD} and \ref{AD2}}\label{PADloc}
Only the proof of Theorem \ref{AD} will be carried out explicitly. Theorem \ref{AD2} can be proven by using the same arguments as in the proof of Theorem \ref{AD}.\\
To accomplish the proof of part (a) we adopt the method of \cite{AH2004} or \citet{HKPS2011}, that is we find a sequence $N=N(m,x)$ such that:
\begeq
P(\tp>N) = P\left(\max\limits_{1\leq k \leq N} \frac{\Qpil}{\hat{\sigma}_m c^{\page}_{1,\alpha}\,\gtilde}\leq 1\right) \limarm \overline{\Psi}(x)\quad\text{for all real }x.
\eneq
To ease notation we will write $a_m = a_m(c^{\page}_{1,\alpha})$, $b_m = b_m(c^{\page}_{1,\alpha})$ and $c=c^{\page}_{1,\alpha}$. As can be seen in the proofs the sequence $N(m,x)$ can be chosen as
\begeq
N = N(m,x) = \left(\frac{\sigma\,c\,m^{\oh-\g}}{\Delta_m} +\frac{\kb}{\am^\g} - \sigma x \frac{\am^{\oh - \g}(1-\g)}{\Dm(1-\g(1-\frac{\kb}{\am}))}\right)^{1/(1-\g)}\mspace{-55mu}.\label{N-def}
\eneq
Before we start with the proof of Theorem \ref{AD} we give some facts that will be useful in the proofs:
\begin{re}\label{re2} 
It is obvious that $\Qm{k}$ can be rewritten as
 \begin{align}
\Qm{k} &= \sum_{i=m+1}^{m+k}\eps_i -\frac{k}{m}\sum_{i = 1}^m \eps_i + \Delta_m(k-\kb+1)I_{\{k\geq \kb\}}\quad\text{for }k=1,2,\ldots,\label{Qmk}
\end{align}
and consequently with $\emb = \ofrac{m}\sum_{\ell=1}^m\eps_\ell$ we have
\begin{align}
\Qpil &\leq \abs{\sume{i}{k}}+k\abs{\emb}+\abs{\Dm(k-\kb +1)I_{\{k\geq\kb\}}}+\abs{\minnik\Qm{i}}\label{re2-eq1}\\
\intertext{and}
\frac{\abs{\minnik\Qm{i}}}{\gtilde} &\leq \maxnik\frac{\abs{\sume{j}{i}}}{\gtildemi}+\maxnik \frac{i\abs{\emb}}{\gtildemi} + \maxnik\frac{\abs{\Dm(i-\kb +1)I_{\{i\geq\kb\}}}}{\gtildemi}\label{re2-eq2}.
\end{align}
\end{re}
\begin{prop}\label{prop1}Introducing the notation $r_m \approx s_m$ for $r_m = s_m(1+ o(1))$ we get under the assumptions of Theorem \ref{AD} that
\begeqO
\am \approx \begin{cases}
             \bfrac{\sigma\, c\,m^{\oh-\g}}{\Dm}^\ofrac{1-\g},& \text{under \eqref{I}},\\
             \quad d_2\,\kb,& \text{under \eqref{II} with }d_2 = \lr{\frac{\sigma\,c}{c_1} + d_1^\g}^\ofrac{1-\g}\text{ and}\\
             \quad\;\kb,& \text{under \eqref{III}.}
            \end{cases}
\eneqO
\end{prop}
\pr From the definition of $\am$ it follows obviously that this definition is equivalent to 
\begeq
\am = \frac{\sigma\,c\,m^{\oh-\g}}{\Dm}\am^\g + \kb \label{prop1-eq-def-am}
\eneq
which yields $\am\geq \kb$ and thus $\frac\kb\am\leq 1$. Now under \eqref{I} the assertion follows directly by
\begeqO
\am = \frac{\sigma\,c\,m^{\oh-\g}}{\Delta_m}\am^\g\lr{1+\frac{\Dm \kb^{1-\g}}{\sigma\,c\,m^{\oh-\g}}\bfrac{\kb}{\am}^\g} = \bfrac{\sigma\,c\,m^{\oh-\g}}{\Delta_m}^\ofrac{1-\g}(1+\lilo{1}).
\eneqO
Under \eqref{II} we have $\limm \kb/\am = d_1$ (cf. Lemma \ref{L1} a) (iv)) and with \eqref{re1-eq2} consequently
\begeqO
\am = \kb\lr{\sigma\,c\,\Dm^{-1}m^{\oh-\g} \kb^{\g-1} + \bfrac{\kb}{\am}^\g}^\ofrac{1-\g}\approx \kb\lr{\frac{\sigma\,c}{c_1} + d_1^\g}^\ofrac{1-\g} = d_2\kb 
\eneqO
Finally under \eqref{III} we have from the definition of $\am$
\begeqO
\frac{\am}{\kb} = \lr{\bfrac{\kb}{\am}^\g + \sigma\,c\,\Dm^{-1} m^{\oh-\g}\kb^{\g-1}}^\ofrac{1-\g} = \bigo{1},
\eneqO
leading to
\begeqO
\am^{1-\g} = \frac{\kb}{\am^\g}\lr{1+\frac{\sigma\,c\,m^{\oh-\g}}{\Delta_m \kb^{1-\g}}\bfrac{\am}{\kb}^\g} = \frac{\kb}{\am^\g}\lr{1+\lilo{1}},
\eneqO
and thus concluding the proof.\prendwol                  
\begin{Le}\label{L1} Let $\g\in[0,\oh)$ and let \eqref{A3} -- \eqref{A5} be satisfied. Then
 \begin{enumerate}[a)]
  \item 
  \begin{enumerate}[(i)]
  \item $\am/m\longrightarrow 0,$
  \item $\sqrt{\am} \Delta_m\longrightarrow\infty,$
  \item $\kb/m\longrightarrow 0,$\label{L1aiv}
  \item $\kb/\am\longrightarrow \begin{cases} 0,&\text{under \eqref{I},}\\
                                         d_1\in(0,1),&\text{under \eqref{II} and}\\
                                         1,&\text{under \eqref{III}.}
                                        \end{cases}
$\label{L1aiii}
\end{enumerate}
\item $N/\am\lra 1$ and consequently the statements of part a) still hold after substitution of $\am$ by $N$.
\item Furthermore
\begeq\limm \ofrac{\sigma}\bfrac{N}{m}^{\g-\oh}\left(\sigma\,c - \frac{\Delta_m}{m^{\oh-\g}}\lr{N^{1-\g}-\frac{\kb}{N^\g}}\right) = x\quad\text{for all real }x.\label{L1.4c}
\eneq
\end{enumerate}
\end{Le}
\pr
a)~Part (iii) follows directly from Assumption \eqref{A1}.\\
(i)~Under \eqref{I} with Proposition \ref{prop1} and Assumption \eqref{A4}
  \[
    \frac{\am}{m}\approx\bfrac{\sigma\,c}{\sqrt{m}\Dm}^\ofrac{1-\g}\limarm 0,
  \]
  under \eqref{II} and \eqref{III} we get the result again with Proposition \ref{prop1} and part (iii) of this Lemma from
  \[
   \frac{\am}{m}\approx\begin{cases}
                        d_2 \kb/m,&\text{under \eqref{II} and}\\
                        \kb/m&\text{under \eqref{III}.}
                       \end{cases}
  \]
(ii)~Because $\am\geq \lr{\sigma\,c\,\Dm^{-1} m^{\oh-\g}}^\ofrac{1-\g}$ we have by Assumption \eqref{A4}
  \[
   \sqrt{\am}\Dm \geq \lr{\lr{\sigma\,c\,\Dm^{-1} m^{\oh-\g}}\Dm^{2(1-\g)}}^\ofrac{2(1-\g)} = (\sigma\,c)^\ofrac{2(1-\g)}\lr{\sqrt{m}\Dm}^\frac{\oh-\g}{1-\g}\limarm\infty
  \]
(iv)~Under \eqref{I} and \eqref{III} the result follows directly from Proposition \ref{prop1} and its proof, under \eqref{II} consider
\[
 \frac{\am}{\kb} = \frac{\sigma\,c\,m^{\oh-\g}}{\Dm \kb^{1-\g}}\bfrac{\am}{\kb}^\g + 1 = \frac{\sigma\,c}{c_1}\bfrac{\am}{\kb}^\g + 1 + \lilo{1}.
\]
Now it can easily be seen that because of the definition of $\am$ the term $\am/\kb$ solving the equation above converges towards a real number $d_1^{-1}\in(1,\infty)$ and hence $\kb/\am\limarm d_1\in(0,\infty).$ Hence we find $d_1$ as the solution of 
\[d_1 = \limm \kb/\am = 1 - \limm \sigma\,c\,\Dm^{-1}m^{\oh-\g}\kb^{\g-1}\lr{\kb/\am}^{1-\g} = 1- \lr{\sigma\,c/c_1}d_1^{1-\g}.\]

b)~It is enough to consider
\[
 \frac{N^{1-\g}}{\am^{1-\g}} = 1-\sigma x \frac{\am^{\oh-\g}(1-\g)}{\am^{1-\g}\Dm\lr{1-\g\lr{1-\kb/\am}}} = 1-\sigma x\frac{1-\g}{\sqrt{\am}\Dm\lr{1-\g\lr{1-\kb/\am}}}.
\]
But aside from Lemma \ref{L1} a) (ii) giving us $\sqrt{\am}\Dm\limarm \infty$ we have
\[
 1-\g\lr{1-\kb/\am}\limarm\begin{cases}
                                   1-\g>0,&\text{under (I) and \eqref{I},}\\
                                   1-\g(1-d_1)>0,&\text{under \eqref{II},}\\
                                   1,&\text{under \eqref{III},}
                                  \end{cases}
\]
which yields the desired result.\\
c)~To ease the notation we first introduce
\begin{align}
u_\g(s,t) = \frac{1-\g}{1-\g\lr{1-s/t}}.\label{u-def}
\end{align}
By inserting the definition of $N$ in $N^{1-\g}$, from \eqref{N-def}, we get 
\begin{align*}
       & \lr{N/m}^{\g-\oh}\left(c - \Dm m^{\g-\oh}\lr{N^{1-\g}-\kb/N^\g}\right)\\
       =&\frac{\Dm\kb}{\sqrt{N}}\lr{1-\lr{N^{1-\g}/\am^{1-\g}}^\frac{\g}{1-\g}} + \sigma x \lr{\am/N}^{\oh-\g}u_\g(\kb,\am)\\
        = & \frac{\Dm\kb}{\sqrt{N}}\lr{1^\frac{\g}{1-\g}-\lr{1+\frac{N^{1-\g}-\am^{1-\g}}{\am^{1-\g}}}^\frac{\g}{1-\g}} + \sigma x \lr{\am/N}^{\oh-\g}u_\g(\kb,\am)\\
        =&A_{\aindp},
       \intertext{so by the mean value theorem we can find $\xi_m$ between 1 and $\lr{N/\am}^{1-\g}$ (which satisfies $\xi_m\to 1$ because of part b)) such that}
       A_\aind = & \sigma x \lr{\am/N}^{\oh-\g}u_\g(\kb,\am) + \frac{\Dm\kb}{\sqrt{N}\am^{1-\g}}\lr{\sigma x \frac{\am^{\oh-\g}}{\Dm}u_\g(\kb,\am)}\frac{\g}{1-\g}\xi_m^\frac{2\g-1}{1-\g}\\
        = & \sigma x \lr{\am/N}^{\oh-\g}u_\g(\kb,\am)\lr{1 + \frac{\g}{1-\g}\frac{\kb}{\am}\frac{\am^\g}{N^\g}\xi_m^\frac{2\g-1}{1-\g}}\\
        = & \sigma x \bfrac{\am}{N}^{\oh-\g}\lr{1 + \frac{\g}{1-\g}\frac{\kb}{\am}\frac{\am^\g}{N^\g}\xi_m^\frac{2\g-1}{1-\g}}\lr{1 - \frac{\g}{1-\g}\frac{\kb}{\am}}^{-1}\quad\limarm \sigma x,
      \end{align*}
      which completes the proof of Lemma \ref{L1}.\prendwol

To prove Theorem \ref{AD} we formulate a set of lemmas containing stepwise approximations of the detector that finally give us the desired asymptotics. For these first steps we follow again the outline of the proofs in \citet{AH2004}. The first step of the proof is to show that the observations before the change-point do not have an impact on the asymptotics under the alternative.
\begin{Le}\label{L2}
Let $\g\in [0,\oh)$. If \eqref{A1} -- \eqref{A5} 
hold, then
\[
\bfrac{N}{m}^{\g-\oh}\left(\maxkkb\frac{\Qpil}{\gtilde} - \frac{\Delta_m (N-\kb)}{\sqrt{m}\lr{N/m}^\g}\right)\limP -\infty.
\]
\end{Le}
\pr
First we note that $\gtilde = m^{\oh-\g}\lr{1+k/m}^{1-\g}k^\g\geq m^{\oh-\g}k^\g$.
Because the indicator function in \eqref{Qmk} equals zero for $1\leq k<\kb$ and because of \eqref{re2-eq1} it is enough to consider
\begin{align*}
&\bfrac{N}{m}^{\g-\oh}\maxkkb \ofrac{\gtilde}\abs{\sume{i}{k}}+\bfrac{N}{m}^{\g-\oh}\maxkkb \frac{k\abs{\emb}}{\gtilde}\\
&+\bfrac{N}{m}^{\g-\oh}\maxkkb \ofrac{\gtilde} \abs{\minnik\Qm{i}}+ \bfrac{N}{m}^{\g-\oh} \frac{\Delta_m (N-\kb)}{\sqrt{m}\lr{N/m}^\g}\\
=&~A_{\aindp}+A_{\aindp}+A_{\aindp}+A_{\aindp}
\end{align*}
We will first show that all but the deterministic term $A_{\aind}$ are stochastically bounded and therefore they do not contribute to the asymptotics. Then it is sufficient to show the divergence of $A_{\aind}$ to prove the lemma. We begin with the term $A_2$ and replace the partial sum of the error terms by a Wiener process and have with Lemma \ref{L1} a)(iv) and b)
\begin{align*}
 \bfrac{N}{m}^{\g-\oh}\maxkkb \ofrac{\gtilde} \abs{\sume{i}{k} - \sigma W_m(k)} &= \Op\max\limits_{1\leq k<\kb} \frac{k^{1/\nu}}{N^{\oh-\g}k^\g}\\
 &= \Opo{\lr{\kb/N}^{\oh-\g}}\\
 &= \Op.
 \end{align*}
We note that
\begin{align*}
 \bfrac{N}{m}^{\g-\oh}\maxkkb \frac{W_m(k)}{\gtilde} \leq \sup\limits_{0< t \leq\kb} \frac{W_m(t)}{\sqrt{N}\lr{t/N}^\g} \eqD \sup\limits_{0< t \leq\kb/N} \frac{W(t)}{t^\g} &= \Op,
\end{align*}
where the equality in distribution comes from the scaling property of the Wiener process.\\
For $A_3$ Lemma \ref{L1} a) (iii), (iv) and Assumption \eqref{A1} yield
\begin{align*}
 \bfrac{N}{m}^{\g-\oh}\maxkkb \frac{k\abs{\emb}}{\gtilde}
 &\leq \frac{\kb^{1-\g}}{m{N^{\oh-\g}}}\abs{\sumem}= \Opo{\lr{\kb/N}^{\oh-\g}\lr{\kb/m}^{\oh}} = \op.
\end{align*}
For $A_4$ it follows by \eqref{re2-eq2} that
\begin{align*}
\maxkkb \ofrac{\gtilde}\abs{\minnik\Qm{i}} \leq \maxkkb \ofrac{\gtilde}\abs{\sume{j}{k}}+\maxkkb \frac{k\abs{\emb}}{\gtilde}=\Op.
\end{align*}
Thus we only have to consider the deterministic term 
\begeqO
A_5=\bfrac{N}{m}^{\g-\oh}\frac{\Dm (N-\kb)}{\sqrt{m}\lr{N/m}^\g} = \frac{\Dm}{N^{\oh-\g}}\lr{N^{1-\g} - \frac{\kb}{N^\g}} = \Dm\sqrt{N}\lr{1-\frac{\kb}{N}}.
\eneqO
It is obvious that the right hand side under \eqref{I} and \eqref{II} tends to infinity. Under \eqref{III} we have with $u_\g$ from \eqref{u-def} that
\begin{align*}
 \frac{\Dm}{N^{\oh-\g}}\lr{
\frac{N-\kb}{N^\g} 
 } &= \frac{\Dm}{N^{\oh-\g}}\lr{\frac{\sigma\,c\,m^{\oh-\g}}{\Dm} - \sigma x \frac{\am^{\oh-\g}}{\Dm}u_\g(\kb,N)
 + \frac{\kb}{N^\g}\lr{\bfrac{N}{\am}^\g - 1}}\\
 & = \sigma\,c\bfrac{m}{N}^{\oh-\g} - \sigma x \bfrac{\am}{N}^{\oh-\g}u_\g(\kb,N)
 + \frac{\Dm\kb}{\sqrt{N}}\lr{\bfrac{N}{\am}^\g - 1}.
\end{align*}
From Lemma \ref{L1} a) (i) it follows directly that the first term diverges, i.e., $\sigma\,c\lr{m/N}^{\oh-\g}\limarm\infty$, for the second term by Lemma \ref{L1} a) (iv) and b) it is clear that this term is bounded.\\
The third term can be treated analogously to the proof of Lemma \ref{L1} c) applying the mean value theorem:
\begin{align*}
 &\frac{\Dm\kb}{\sqrt{N}}\lr{\lr{1+\frac{N^{1-\g}-\am^{1-\g}}{\am^{1-\g}}}^{\g/(1-\g)} - 1}\\
 =&\frac{\Dm\kb}{\sqrt{N}\am^{1-\g}}\lr{-\sigma x \frac{\am^{\oh-\g}}{\Dm}u_\g(\kb,N)}\frac{\g}{1-\g}\xi_m^{({2\g-1})/({1-\g})}\\
 =&-\sigma x\,u_\g(\kb,N) \frac{\kb}{\sqrt{N\am}}\xi_m^{({2\g-1})/({1-\g})}\\
 =& \bigo{1}.
\end{align*}
This gives us the desired result.
\prendwol

The next step is an approximation of our detector by functionals of a sequence of Wiener processes. To ease the notation we define
\begin{align}
\wdr{j} &= \sigma W_m(j) + (j-\kb + 1)\Dm\ind{j\geq\kb}\label{def-WD}
\intertext{and}
\pw{m}{k} &= \wdr{k} - \minnik\wdr{i}.\label{def-PW}
\end{align}
\begin{Le}\label{L3}
Let $\g\in [0,\oh)$ and Assumptions \eqref{A1} -- \eqref{A5} hold.
Then
\begin{equation*}
\vf \maxkbkN \ofrac{\gtilde} \abs{\Qpil - \pw{m}{k}} = \op.
\end{equation*}
\end{Le}
\pr
The deterministic terms cancel out thus using
\begin{align*}
& \maxkbkN\ofrac{\gtilde}\abs{\minnik \wdr{i} - \minnik \Qm{i}}\\
\leq & \maxkbkN\ofrac{\gtilde}\maxnik\abs{ \sigma W_m(i)- \lr{\sume{j}{i} - \frac im \sumem}}\\
\leq & \maxkN\ofrac{\gtilde}\lr{\abs{\sume{j}{k} - \sigma W_m(k)} + \frac km \abs{\sumem}}
\end{align*}
it is sufficient to consider
\begin{align*}
& \vf \maxkN\ofrac{\gtilde}\abs{ \sume{j}{k} - \sigma W_m(k)}
=
A_{\aindp},
\intertext{for which with Assumption \eqref{A2} and because $y^{1/\nu-\g}$ is monotone we have}
A_{\aind}= & \Op\maxkN\frac{k^{1/\nu}}{\VGFN} = \Op N^{\g-\oh}\max\left\{1,N^{1/\nu-\g}\right\}\\
= & \Op \max\left\{N^{\g-\oh},N^{1/\nu-\oh}\right\} = \Op\lilo{1}\\
= & \op, 
\end{align*}
and
\begin{align*}
 \vf \maxkN \left.{\frac km \abs{\sumem}}\right/{\gtilde} & \leq \vf \frac 1m\abs{\sumem}\maxkN\frac{k^{1-\g}}{m^{\oh-\g}}\\
 & = \frac{\sqrt{N}}{m}\Opo{\sqrt{m}} = \Opo{\sqrt{N/m}}\\
 & = \op.&
\end{align*}
\prend
The boundary function $\gtilde$ can be replaced by an asymptotically equivalent function that simplifies the coming calculations.
\begin{Le}\label{L4}
Let $\g\in [0,\oh)$ and Assumptions \eqref{A1} -- \eqref{A5} hold and define
\begin{equation*}
h(m,k) = \abs{\ofrac{\gtilde} - \ofrac{\sbfk}}. 
\end{equation*}
Then
\begin{equation*}
\vf \maxkbkN \abs{\pw{m}{k}}h(m,k) = \op.
\end{equation*}
\end{Le}
\pr
\citet{AH2004} showed that 
\begeqO
\vf\maxkbkN \sigma\abs{W_m(k)} h(m,k) = \op.
\eneqO
For the deterministic term $(k-\kb+1)\Dm$ we get 
\begin{align}
&\vf\maxkbkN \Dm (k-\kb+1) h(m,k) \notag\\
= & \frac{\Dm}{N^{\oh-\g}} \maxkbkN \frac{k-\kb+1}{k^\g}\left(1^{1-\g} - \bfrac{m}{k+m}^{1-\g}\right)\notag\\
=& \frac{\Dm}{N^{\oh-\g}}\maxkbkN k^{1-\g}\left(1^{1-\g} - \bfrac{m}{k+m}^{1-\g}\right)\notag\\
=&A_{\aindp}.\notag
\end{align}
By application of the mean value theorem for every $\kb\leq k\leq N$ we can find a real number $\xi_{m,k}$ satisfying $m/(m+k) < \xi_{m,k} < 1$ such that
\begin{align*}
A_\aind =
 & \frac{\Dm}{N^{\oh-\g}} \maxkbkN \frac{k-\kb+1}{k^\g}\frac{k}{k+m}(1-\g)\xi_{m,k}^{-\g}.\\
 \intertext{Because $\xi_{m,k}^{-\g}\leq \lr{m/(m+k)}^{-\g}$ and $(k-\kb+1)\lr{k/(k+m)}^{1-\g}$ is strictly increasing in $k$ we have}
A_\aind \leq&\frac{\Dm}{N^{\oh-\g}m^\g}\maxkbkN (k-\kb+1)\bfrac{k}{k+m}\\
=& \frac{\Dm}{N^{\oh-\g}}(N-\kb+1)\bfrac{N}{N+m}^{1-\g}\\
\leq & \frac{\Dm \sqrt{N}(N-\kb+1)}{m} \\
=& \frac{\Dm \sqrt{N}(N-\kb)}{m} + \frac{\Dm \sqrt{N}}{m}.
\end{align*}
Here the last term clearly tends to 0 and for the first term we get by Lemma \ref{L1} b) and from \eqref{prop1-eq-def-am}
\begin{align*}
\frac{\Dm \sqrt{N}(N-\kb)}{m} \approx & \frac{\Dm \sqrt{\am}(\am-\kb)}{m} = \frac{\Dm \sqrt{\am}}{m}\frac{\sigma\,c\,m^{\oh-\g}\am^\g}{\Dm} = \sigma\,c\bfrac{\am}{m}^{\oh+\g} = o(1),
\end{align*}
where the last equality comes from Lemma \ref{L1} a) (i). For the minimum term the claim follows with the same arguments and this is completing the proof.
\prend
Before we can state the next lemma we again have to introduce some notation. For $\delta\in(0,1)$ we define 
\[\Nd = (1-\delta)N, \quad \Nb = N - \kb -1\quad\text{and}\quad \Nbd = (1-\delta)\Nb.\] 
\begin{Le}\label{L5}
 Let $\g\in[0,\oh)$ and \eqref{A1} -- \eqref{A5} hold. Then for every $\delta\in(0,1)$
 \begin{enumerate}[a)]
  \item $\limm P\lr{\maxkbkN\frac{\pw{m}{k}}{\sbfk} = \maxdelkN \frac{\pw{m}{k}}{\sbfk}} = 1,$
  \item $\limm P\lr{\maxkbkNb\frac{\sigma W_{m}(j) + \Dm j}{\sqrt{N}} = \maxdelkNb \frac{\sigma W_{m}(j) + \Dm j}{\sqrt{N}}} = 1.$
 \end{enumerate}
\end{Le}
\pr
a) We note (cf. \citet{AH2004}) that
\begeq
\maxkbkN\frac{\abs{W(k)}}{\sbfk} = \Opo{\bfrac{N}{m}^{\oh-\g}} = \opo{\frac{\Dm N^{1-\g}}{m^{\oh-\g}}}.\label{***}
\eneq
Now it can be seen easily that this result also holds true for the extended range $0\leq k\leq N$.\\
Then
\begin{align*}
 &\;P\left(\maxkbkN\frac{\pw{m}{k}}{\sbfk} > \maxdelkN \frac{\pw{m}{k}}{\sbfk}\right)\\
  =&\;P\left(\bigcup_{k = \kb}^{\floor{\Nd}}\bigcap_{\ell = \Nd}^N\left\{k^{-\g}\pw{m}{k} > \ell^{-\g}\pw{m}{\ell}\right\}\right)\\
 \leq &\; P\left(\bigcup_{k = \kb}^{\floor{\Nd}}\left\{k^{-\g}\pw{m}{k} > N^{-\g}\pw{m}{N}\right\}\right).
\end{align*}
We can rewrite
\[
 k^{-\g}\pw{m}{k} > N^{-\g}\pw{m}{N}
\]
as
\begin{align}
\kNmg \lr{ \wdr{k} - \minnik \wdr{i}} -\sigma W_m(N) + \minniN\wdr{i}  &> \Dm (N - \kb +1),\label{Pr-L5}
\end{align}
where the term on the right can be replaced by $\Dm N$ to get an upper bound for the probability of \eqref{Pr-L5}.
Because for $k\in[\kb,\Nd)$
\begin{align*}
\bfrac{k}{N}^{-\g}\minnik \lr{\sigma W_m(i) + (i-\kb+1)\Dm I_{\{i\geq\kb\}}} &\geq \minnik \frac{\sigma W_m(i)}{\lr{i/N}^\g}\geq - \maxniN \frac{\sigma (-W_m(i))}{\lr{i/N}^\g},
\end{align*}
it follows that
\begeqO
\relphantom{\leq\;}P\left(\bigcup_{k = \kb}^{\floor{\Nd}}\left\{k^{-\g}\pw{m}{k} > N^{-\g}\pw{m}{N}\right\}\right)
\eneqO
\begeqO
  \leq\; P \Biggl(\maxkbkdelN\left(\frac{\sigma W_m(k)}{\kNg\Dm N}\right) &+& \maxkbkdelN\left(\frac{k-\kb+1}{\kNg N}\right) \\
  &+&
   \maxniN\left(\frac{\sigma (-W_m(i))}{\Dm N\lr{i/N}^\g}\right)- \sigma \frac{W_m(N)}{\Dm N} > 1 \Biggr).
\eneqO
Now \eqref{***} yields
\begin{align*}
\maxkbkdelN \frac{\sigma W_m(k)}{\kNg\Dm N} &= \op,\\
\maxniN \frac{\sigma W_m(i)}{\Dm N\lr{i/N}^\g} &= \op,\\
\sigma\frac{W_m(N)}{\Dm N} &= \op.
\end{align*}
Thus with
\begin{equation*}
\maxkbkdelN \frac{k-\kb+1}{\kNg N} \leq \frac{\Nd -\kb + 1}{\lr{\Nd/N}^\g N} = (1-\delta)^{1-\g} - \frac{\kb}{(1-\delta)^\g N} + \ofrac{(1-\delta)^\g N} <1,
\end{equation*}
for large enough $m$, we have
\begeqO
  P \Biggl(\maxkbkdelN\left(\frac{\sigma W_m(k)}{\kNg\Dm N}\right) &+& \maxkbkdelN\left(\frac{k-\kb+1}{\kNg N}\right) \\
  &+&
   \maxniN\left(\frac{\sigma (-W_m(i))}{\Dm N\lr{i/N}^\g}\right)- \sigma \frac{W_m(N)}{\Dm N} > 1 \Biggr)\limarm 0.
\eneqO
b) Similarly we get
\begin{align*}
&P\lr{\maxkbkNb\frac{\sigma W_{m}(j) + \Dm j}{\sqrt{N}} > \maxdelkNb \frac{\sigma W_{m}(j) + \Dm j}{\sqrt{N}}}\\
\leq\;& P\lr{\bigcup_{j = \kb}^{\floor{\Nbd}}\left\{\sigma W_m(j) + \Dm j > \sigma W_m(\Nb) + \Dm \Nb\right\}}\\
\leq\;& P\lr{\maxkbkdelNb \frac{\sigma W_m(j)}{\Dm \Nb} + \maxkbkdelNb \frac{j}{\Nb} - \frac{\sigma W_m(\Nb)}{\sqrt{\Nb}}\ofrac{\Dm\sqrt{\Nb}} > 1}.
\end{align*}

We first consider the term $\Dm\sqrt{\Nb} = \Dm \sqrt{N}\sqrt{1-\lr{\kb+1}/{N}}$. Under \eqref{I} and \eqref{II} it is obvious that $\Dm\sqrt{\Nb}\limarm\infty$, under \eqref{III} with Lemma \ref{L1} b) and \eqref{prop1-eq-def-am}
\[
 \Dm\sqrt{\Nb}\approx \Dm\sqrt{\am - \kb - 1} = \Dm\sqrt{\frac{\sigma\,c\,m^{\oh-\g}}{\Dm}\am^\g - 1} = \sqrt{\sigma\,c\,\Dm m^{\oh}\bfrac{\am}{m}^\g - \Dm^2}.
\]
Because of Assumption \eqref{A3} it is enough to consider
\[
 \Dm m^{\oh}\lr{\am/m}^\g \approx \Dm m^{\oh-\g+\g\beta} = \Dm m^{\beta(1-\g) - \oh +\g}m^{2(\g\beta + \oh -\g) - \beta},
\]
but since $2(\g\beta + \oh -\g) - \beta \geq 0$ under $\g<1/2$ is equivalent to $\beta\leq 1$ we have $\Dm\sqrt{\Nb}\limarm\infty$. 
This together with $\Nb^{-\oh}W_m(\Nb) = \Op$ gives us
\begin{equation*}
\frac{W_m(\Nb)}{\sqrt{\Nb}}\ofrac{\Dm\sqrt{\Nb}} = \op.
\end{equation*}
The rest of the proof follows analogously to part a) of this proof.\prendwol
\begin{Le}\label{L6}
Let $\g\in[0,\oh)$. If \eqref{A1} -- \eqref{A5} are satisfied, then 
\begin{equation}
P\lr{\maxkbkN \frac{\pw{m}{k}}{\sbfk} \leq \sigma\,c}\limarm \overline{\Psi}(x)\quad\text{for all real }x.\notag
\end{equation}
\end{Le}
\pr
Application of Lemma \ref{L5} a) and then letting $\delta\downarrow 0$ yields
\begin{align}
 &\limm P\lr{\maxkbkN \frac{\pw{m}{k}}{\sbfk} \leq \sigma\,c}\notag\\
 =& \limm P\lr{\maxdelkN \frac{\pw{m}{k}}{\sbfk} \leq \sigma\,c}\notag\\
=& \limm P\lr{\ofrac{\sqrt{N}}\lr{\wdr{N} - \minniN \wdr{i}}\leq \sigma\,c\vf}.\label{L6-p1}
\end{align}
But by replacing $N-i$ with $j$ and denoting $\widetilde{N} = N-\kb + 1$ we get
\begin{align*}
 &\ofrac{\sqrt{N}}\lr{\wdr{N} - \minniN \wdr{i}}\\
 =&\maxniN\ofrac{\sqrt{N}}\lr{\sigma(W_m(N)-W_m(i)) + \Dm\lr{\widetilde{N} - (i-\kb +1)\ind{i\geq\kb}}}\\
 =&\maxnjN\ofrac{\sqrt{N}}\lr{\sigma(W_m(N)-W_m(N-j)) + \Dm\lr{\widetilde{N} - (\widetilde{N}-j)\ind{N-\kb\geq j}}}\\
 =&\max\Biggl\{\hspace{2mm}\max_{0\leq j\leq N-\kb}\ofrac{\sqrt{N}}\lr{\sigma(W_m(N)-W_m(N-j)) + \Dm j},\\
 &\relphantom{\max\Biggl\{}\max_{N-\kb< j\leq N}\ofrac{\sqrt{N}}\lr{\sigma(W_m(N)-W_m(N-j)) + \Dm \widetilde{N}}\Biggr\}.\\
 \intertext{Because of the time reversibility of the Wiener process (cf. \citet{borodin2002}) we can find a sequence of Wiener processes $\{W_{1,m}(t), t\geq 0\}_{m=1,2,\ldots}$ such that}
 &\max\Biggl\{\hspace{2mm}\max_{0\leq j\leq N-\kb}\ofrac{\sqrt{N}}\lr{\sigma(W_m(N)-W_m(N-j)) + \Dm j},\\
 &\relphantom{\max\Biggl\{}\max_{N-\kb< j\leq N}\ofrac{\sqrt{N}}\lr{\sigma(W_m(N)-W_m(N-j)) + \Dm \widetilde{N}}\Biggr\}\\
 \eqD & \max\Biggl\{\hspace{2mm}\max_{0\leq j\leq N-\kb}\ofrac{\sqrt{N}}\lr{\sigma W_{1,m}(j) + \Dm j},\\
 &\relphantom{\max\Biggl\{}\max_{N-\kb < j\leq N}\ofrac{\sqrt{N}}\lr{\sigma W_{1,m}(j) + \Dm \widetilde{N}}\Biggr\}.
\end{align*}
Applying Lemma \ref{L5} b) and again letting $\delta\downarrow 0$ we see that the first term in the outer maximum is taking its maximum arbitraryly close to $N-\kb$ and hence can be omitted and we can proceed with \eqref{L6-p1} to have
\begin{align}
& \limm P\lr{\ofrac{\sqrt{N}}\lr{\wdr{N} - \minniN \wdr{i}}\leq \sigma\,c\vf}\notag\\
=& \limm P\lr{\max_{N-\kb\leq j\leq N}\ofrac{\sqrt{N}}\lr{\sigma W_{1,m}(j) + \Dm \widetilde{N}}\leq \sigma\,c\vf}\notag\\
=& A_{\aindp}\notag
\intertext{Due to the scaling property of the Wiener process again we can find a sequence $\{W_{2,m}(t), t\geq 0\}_{m=1,2,\ldots}$ of Wiener processes such that}
A_{\aind}=& \limm P\lr{\max_{1-\frac{\kb}{N}\leq \frac{j}{N}\leq 1} W_{2,m}(j/N) \leq \ofrac{\sigma}\vf\lr{\sigma\,c - \frac{\Dm (N-\kb)}{m^{\oh-\g}N^\g}}}\notag\\
= & \overline{\Psi}(x),\notag
\end{align}
where the last equation follows from Lemma \ref{L1} a) (iv), c) and Slutsky's Lemma.
\prend
\\[1mm]
\textbf{Proof of Theorem \ref{AD} (a):}\\[3mm] 
The major part of the proof is just a combination of the preceding lemmas. With $u_\g$ from \eqref{u-def} the rest follows along the lines of the proof of Theorem 1.1 of \cite{AH2004}:
\begin{align*}
 \Psi(x) &= 1 - \overline{\Psi}(-x) \\
 &= 1-\limm P\lr{\tp > N(m,-x)}\\
&= \limm P\lr{\tp \leq N(m,-x)}\\
&= \limm P\lr{\lr{\tp}^{1-\g} \leq \lr{N(m,-x)}^{1-\g}}\\
 &= \limm P\lr{\lr{\tp}^{1-\g} - \am^{1-\g} \leq  \sigma x \frac{\am^{\oh-\g}}{\Dm}u_\g(\kb,\am)}\\
&= \limm P\lr{\lr{\tp}^{1-\g} - \am^{1-\g} \leq  \sigma x \frac{\am^{\oh-\g}}{\Dm}u_\g(\kb,\am)}\\
&= \limm P\lr{\frac{\am^\g}{1-\g}\frac{\lr{\tp}^{1-\g} - \am^{1-\g}}{\bm} \leq  \sigma x \frac{\sqrt{\am}}{\Dm\lr{1-\g\lr{1-\frac\kb\am}}}\bm^{-1}}\\
&= \limm P\lr{\frac{\tp - \am}{\bm} \leq x},
\end{align*}
where the last equation follows because with the same arguments as in \citet{AH2004} it can be shown that 
$$\frac{\tp - \am}{\bm}\quad\text{ and }\quad\frac{\am^\g}{1-\g}\frac{\lr{\tp}^{1-\g} - \am^{1-\g}}{\bm}$$
 have the same limit distribution.
\prendwol\\[3mm]
\textbf{Proof of Theorem \ref{AD} (b):}\\[3mm]
The proof of part (b) of Theorem \ref{AD} follows analogously to the steps in part (a). In the corresponding step to \eqref{L6-p1}, instead we find
\begin{align*}
&\limm P\lr{\ofrac{\sqrt{N}}\wdr{N}\leq \sigma\,c^Q_{1,\alpha}\vf} \\
=&\limm P\lr{ \ofrac{\sqrt{N}}W_{m}(N) \leq \ofrac{\sigma}\vf\lr{\sigma\,c^Q_{1,\alpha} - \frac{\Dm (N-\kb)}{m^{\oh-\g}N^\g}}}\notag\\
=&\Phi(x),
\end{align*}
which again with the same arguments used in part (a) implies the result.\prendwol
\\[3mm]
\textbf{\large Acknowledgements. }
I thank Alexander Aue, Lajos Horv{\'a}th, and Josef G. Steinebach for the fruitful discussions throughout the work on this paper.
\newpage
\bibliographystyle{plainnat}

\end{document}